\documentclass{article}

\usepackage{caption}
\usepackage{subcaption}
\usepackage{amssymb}
\usepackage{graphicx}
\usepackage{dcolumn}
\usepackage{bm}
\usepackage{amsmath}

\begin{document}

\title{\bf Late time cosmic acceleration from natural infrared cutoff}
\author{Mohammad Ali Gorji\thanks{{email: m.gorji@stu.umz.ac.ir}}\\\\
{\small{\it Department of Physics, Faculty of Basic Sciences,
University of Mazandaran,}}\\{\small{\it P. O. Box 47416-95447,
Babolsar, Iran}}}

\maketitle
\begin{abstract}
In this paper, inspired by the ultraviolet deformation of the 
Friedmann-Lema\^{\i}tre-Robertson-Walker geometry in loop quantum 
cosmology, we formulate an infrared-modified cosmological model. 
We obtain the associated deformed Friedmann and Raychaudhuri 
equations and we show that the late time cosmic acceleration can 
be addressed by the infrared corrections. As a particular example, 
we applied the setup to the case of matter dominated universe. 
This model has the same number of parameters as $\Lambda$CDM, but 
a dynamical dark energy generates in the matter dominated era at 
the late time. According to our model, as the universe expands, 
the energy density of the cold dark matter dilutes and when the 
Hubble parameter approaches to its minimum, the infrared effects 
dominate such that the effective equation of state parameter 
smoothly changes from $w_{_{\rm eff}}=0$ to $w_{_{\rm eff}}=-2$. 
Interestingly and nontrivially, the unstable de Sitter phase 
with $w_{_{\rm eff}}=-1$ is corresponding to $\Omega_m=\Omega_d
=0.5$ and the universe crosses the phantom divide from the 
quintessence phase with $w_{_{\rm eff}}>-1$ and $\Omega_m>
\Omega_d$ to the phantom phase with $w_{_{\rm eff}}<-1$ and $
\Omega_m<\Omega_d$ which shows that the model is observationally 
viable. The results show that the universe finally ends up in a 
big rip singularity for a finite time proportional to the 
inverse of the minimum of the Hubble parameter. Moreover, we 
consider the dynamical stability of the model and we show that 
the universe starts from the matter dominated era at the past 
attractor with $w_{_{\rm eff}}=0$ and ends up in a future 
attractor at the big rip with $w_{_{\rm eff}}=-2$.\vspace{5mm}
\end{abstract}

\section{Introduction}
Cosmological observations indicate that the Universe accelerates positively at the small
redshifts \cite{DE} which leads to the so-called dark energy problem \cite{DEP,DEP2}. In
the standard $\Lambda$CDM model, cosmological constant dominates at the late time and
derives cosmic speed-up. But, the models in favor of cosmological constant clue to the
cosmological constant problem due to the possible identification of the cosmological
constant with the vacuum energy of the quantum fields \cite{DEP,CCP}. Furthermore,
increasing evidences from the cosmological data reveal that the energy density
corresponds to the dark energy evolves very slowly in time and the associated equation
of state parameter lies in a narrow strip around $w=-1$ \cite{DE}. Thus, cosmological
constant with sharp value $w=-1$ for the equation of state parameter is an appropriate
candidate in the first order of approximation \cite{DE-CC}. In order to explain the
dynamical nature of the dark energy, the quintessence scenarios with $w>-1$ and phantom
models with $w<-1$ are proposed. In this respect, one usually interested in models
which support the transition from the quintessence era to the phantom phase. These
scenarios are usually based on two postulates: i) assuming general relativity is
applicable even on cosmological scales and then considering some sort of unusual matter
component(s) costing violation of some energy conditions, ii) deformation of general
relativity at the cosmological scales. For the first case the matter source is usually
given by a scalar field \cite{DE-MG,chaplygin,cardassian} and for the latter case,
there are many candidates such as the extra dimensions models , $f(R)$ theories
\cite{DE-FR,DE-FR2} and recently proposed massive gravity models \cite{DE-NIR}.

From the theoretical point of view, de Sitter spacetime is a maximally symmetric space
and its constant curvature is completely determined by the cosmological constant. Apart 
from the very small variation of cosmological constant with time, it can be interpreted 
as a fundamental constant of nature much similar to the speed of light and Planck 
constant. It therefore provides a universal infrared (IR) cutoff (corresponding to the 
large length scale $\sim10^{-56}\mbox{cm}^{-2}$) for the universe. For instance, 
existence of cosmological constant as an IR cutoff is essential for the quantization of 
scalar field in de Sitter spacetime. More precisely, it provides a 
minimum scale for the momenta of modes through the uncertainty principle and removes the 
IR divergences in this setup \cite{maggiore}. In this respect the uncertainty principle 
will be modified in curved spacetimes in order to respect the existence of cosmological 
constant as a universal IR cutoff \cite{UR-dS}. On the other hand, existence of a minimal 
length scale is suggested by any quantum theory of gravity such as loop quantum gravity 
\cite{Loop} and string theory \cite{String}. It is also shown that the uncertainty 
principle is modified in the presence of a minimal length scale \cite{GUP}. Thus, the 
uncertainty principle gets modifications in IR and ultraviolet (UV) regimes in order to 
respect the existence of cosmological constant and minimal length scale respectively
\cite{KempfIR}. Taking these universal IR and UV cutoffs into account, the quantum 
field theories turn out to be renormalizable \cite{IRCurve}. Therefore, natural IR and 
UV cutoffs would be emerged in the context of ultimate quantum gravity theory. While 
the existence of a universal IR cutoff is supported by the standard general relativity 
framework through the de Sitter spacetime\footnote{Note that the anti-de Sitter 
spacetime with negative cosmological constant is also an appropriate candidate from the 
theoretical point of view. But it rejects by cosmological observations.}, there is not 
any explanation for the UV cutoff (minimal length scale) in this setup. On the other 
hand, a minimal length scale as a UV cutoff emerges in loop quantum gravity framework 
\cite{LQG} but there is not a well-defined explanation for taking a cosmological 
constant into account in this setup (see however Refs. \cite{LQC-CC} where some 
attempts have done in this direction). In this paper, we follow the UV deformation of 
the Friedmann-Lema\^{\i}tre-Robertson-Walker (FLRW) universe in loop quantum cosmology
and we construct the corresponding IR-deformed case. We show that the late time 
cosmic acceleration arises in this setup which is significantly different from 
the $\Lambda$CDM model such that the universe crosses the phantom divide from 
$w_{_{\rm eff}}>-1$ to $w_{_{\rm eff}}<-1$.

\section{FLRW Universe}
The spatial part of the spatially flat FLRW universe is 3-manifold $M$ with the 
Euclidean isometry group and ${\mathbf R}^3$ topology. One then can fix a 
constant orthonormal triad $e^a_i$ and a co-triad $\omega^i_a$ compatible
with a flat fiducial metric $^{o}q_{ab}$ on $M$. The corresponding
gravitational phase space consists of pairs $(A^i_a,E^a_i)$ on $M$, where
$A^i_a$ is a $SU(2)$ connection and $E^a_i$ is its canonically conjugate
field \cite{Bojowald}. Because of the symmetries of the 3-manifold $M$, all
the information of the phase space variables $(A^i_a,E^a_i)$ are summarized
in two variables $(\beta,V)$ which satisfy canonical Poisson algebra
\begin{equation}\label{SPA}
\{\beta,V\}=\frac{\kappa\gamma}{2}\,,
\end{equation}
on two-dimensional phase space $\Gamma$, where $\kappa=8\pi{G}$ (we work
in unit $c=1$, where $c$ is the speed of light in vacuum) and $\gamma
\approx0.2375$ is the Barbero-Immirizi parameter which is fixed by the
black hole entropy calculations in loop quantum gravity \cite{BI}. These
variable are related to the old geometrodynamics variables as
\begin{eqnarray}\label{Old}
\beta=\gamma\,\frac{\dot{a}}{a},\hspace{1cm}V=a^3\,.
\end{eqnarray}
So, $V$ is the comoving volume and its canonically conjugate variable
$\beta$ is (up to a constant) the Hubble parameter.

Considering a perfect fluid as a source for the matter content, the
associated energy density consisting of non-relativistic and
relativistic matters will be a function of volume as $\rho=\rho(V)$
and the corresponding Hamiltonian function is given by
\begin{equation}\label{Hamiltonian}
{\mathcal H}=-\frac{3}{\kappa\gamma^2}\beta^2V+\rho{V}\,.
\end{equation}
The Hamiltonian system of the FLRW universe in terms of Ashtekar
variables $(\beta,V)$ is therefore defined by the relations
(\ref{SPA}) and (\ref{Hamiltonian}) on two-dimensional phase space
$\Gamma$: The kinematics is defined by the Poisson bracket
(\ref{SPA}) and the dynamical evolution is governed by the
Hamiltonian (\ref{Hamiltonian}). It is easy to show that the
associated Hamilton's equations together with the Hamiltonian
constraint ${\mathcal H}\approx0$ lead to the standard Friedmann
and Raychaudhuri equations.
\subsection{UV-deformed Phase Space}
In loop quantum cosmology scenario however this Hamiltonian system
gets holonomy corrections at the UV regime. At the semiclassical
regime, these UV effects can be taken into account in two equivalent
ways on the corresponding UV-deformed phase space $\Gamma_\lambda$.
One can work in noncanonical chart on $\Gamma_\lambda$ in which the
Poisson bracket (\ref{SPA}) gets UV modification while the
Hamiltonian function (\ref{Hamiltonian}) retains its standard
functional form \cite{Taveras}. Equivalently, one can also work in
canonical chart on $\Gamma_\lambda$ such that the form of Poisson
bracket (\ref{SPA}) remains unchanged and the Hamiltonian function
(\ref{Hamiltonian}) gets modified functional form \cite{CPR}. These
two different representations are related to each other through
the Darboux transformation and lead to the same Friedmann and
Raychaudhuri equations \cite{Poly-C-NC,PUR0}. In this paper we
work in the first picture in which the Poisson bracket gets UV
modification as \cite{Taveras}
\begin{equation}\label{PA-UV}
\{\beta,V\}=\frac{\kappa\gamma}{2}\sqrt{1-\lambda^2\beta^2},
\end{equation}
where $\lambda$ is the UV deformation parameter which is preferably
of the order of the Planck length $\lambda\sim{l_{_{\rm Pl}}}$.
Clearly, $\beta$ gets maximum value as $\beta<\lambda^{-1}$ in this
setup. More precisely, the space of $\beta$ is compactified to a
circle $S^1$ with radius $\lambda^{-1}$ \cite{CPR}. The UV-deformed
Poisson algebra (\ref{PA-UV}) implies the following modified
Hamilton's equations
\begin{eqnarray}\label{EoM-UV}
{\dot V}=\{V,{\mathcal H}\}=\frac{\kappa\gamma}{2}\sqrt{1-
\lambda^2\beta^2}\,\,\frac{\partial{\mathcal H}}{\partial\beta}\,,
\end{eqnarray}
\begin{eqnarray}\label{EoM-UV2}
{\dot\beta}=\{\beta,{\mathcal H}\}=-\frac{\kappa\gamma}{2}\sqrt{
1-\lambda^2\beta^2}\,\,\frac{\partial{\mathcal H}}{\partial{V}}\,.
\end{eqnarray}
The above equations correctly reduce to the standard Hamilton's
equations in the limit of $\lambda\rightarrow0$. Substituting
(\ref{Hamiltonian}) into (\ref{EoM-UV}) gives
\begin{equation}\label{Vdot}
{\dot V}=\frac{3V\beta}{\gamma}\sqrt{1-\lambda^2\beta^2}.\nonumber
\end{equation}
Using Hamiltonian constraint ${\mathcal H}\approx0$ and after some
manipulations, it is straightforward to obtain the following
UV-deformed Friedmann equation
\begin{equation}\label{Friedmann-UV}
H^2=\frac{\kappa}{3}\rho\left(1-\frac{\rho}{\rho_{_{\rm max}}}
\right)\,,
\end{equation}
where $H=\frac{\dot a}{a}$ is the Hubble parameter and we have
also defined $\rho_{_{\rm max}}=\frac{3}{\kappa\gamma^2\lambda^2}
$. The energy density and the Hubble parameter get maximum bounds
$\rho\leq\rho_{_{\rm max}}$ and $H<H_{_{\rm max}}=\left(\frac{
\kappa\rho_{_{\rm max}}}{12}\right)^{\frac{1}{2}}$ in this setup
and the big bang singularity problem resolves such that the
initial singularity in standard model of cosmology replaces with
a bounce \cite{LQCBB}. Existence of the minimal length scale as
a UV cutoff for the system thus naturally leads to the spacetime
singularity resolution in cosmological setup. In this paper, we
are interested to study the effect of the existence of an IR
cutoff on the late time cosmic acceleration in order to address
the dark energy problem. As we will show in the next subsection,
taking an IR cutoff into account leads to the self-accelerating
universe at the late time.

\subsection{IR-deformed Phase Space}
In the absence of a fundamental theory at the IR regime, we would
like to construct an IR-deformed Hamiltonian system following the
way by which the UV cutoff is taken into account in the
Hamiltonian system of the FLRW universe in loop quantum cosmology
scenario. In loop quantum cosmology scenario, the space of $\beta
$ is compactified to a circle $S^1$ which leads to the deformed
Friedmann equation (\ref{Friedmann-UV}). For the case of the
IR-deformed Hamiltonian system, we should explore the modification
of the space of $V$. At the first glance, it seems that we should
consider the modified $S^1$ geometry for the space of $V$. But,
let us more elaborate on this point. Indeed, the UV and IR cutoffs
are universal and therefore the space of the associated variable
should be maximally symmetric \cite{NC}. From the global point of
view, for the case of one-dimensional space (with which we are
interested in this paper) there are only two possibilities: a
circle $S^1$ with compact $SO(2)$ symmetry and a hyperbolic space
$H^1$ with open $SO(1,1)$ symmetry . For the UV-deformed phase
space, the unique representation of holonomy-flux algebra fixes
the space of $\beta$ to have $SO(2)$ symmetry which local
coordinatization (\ref{PA-UV}) \cite{Bojowald}. In the absence of
any fundamental theory for the IR sector, we consider both of the
possible symmetries $SO(2)$ and $SO(1,1)$ for the space of $V$.
Furthermore, inspired by the loop quantum cosmology scenario, we
work with Ashtekar variables and also consider the local
coordinatization as same as (\ref{PA-UV}). We therefore lead to
two possible IR-deformed Poisson algebras
\begin{equation}\label{PA-IR}
\{\beta,V\}_{_\pm}=\frac{\kappa\gamma}{2}\sqrt{1\pm\alpha^2V^2},
\end{equation}
where $\alpha$ is the IR deformation parameter with dimension of
inverse of volume. Before applying the above IR-deformed Poisson
algebras to the cosmological setup, let us more elaborate on the
differences between these two IR-deformed models. For the case of
minus sign, clearly there is a maximum value $V_{\rm max}=
\alpha^{-1}$ for the volume of the universe such that $V\in[0,
\alpha^{-1})$ while $V\in[0,\infty)$ for the case of plus sign.
At the quantum level, the model with minus sign should be defined
on a lattice \cite{PUR0,PUR} while the model with plus sign leads
to the generalized uncertainty relation. Following the way
suggested in Ref. \cite{GUP-Kempf}, it is straightforward to show
that the associated uncertainty relation implies a minimum
uncertainty in measurement of the corresponding conjugate
variable $\beta$ (see also Refs. \cite{KempfIR,Mignemi,Battisti}).

Taking the IR-deformed Poisson algebras (\ref{PA-IR}) into account,
the corresponding IR-deformed Hamilton's equations are given by
\begin{eqnarray}\label{EoM-IR}
{\dot V}_{_\pm}=\{V,{\mathcal H}\}_{_\pm}=\frac{\kappa\gamma}{
2}\sqrt{1\pm\alpha^2V^2}\,\,\frac{\partial{\mathcal H}}{
\partial\beta}\,,
\end{eqnarray}
\begin{eqnarray}\label{EoM-IR2}
{\dot\beta}_{_\pm}=\{\beta,{\mathcal H}\}_{_\pm}=-\frac{\kappa
\gamma}{2}\sqrt{1\pm\alpha^2V^2}\,\,\frac{\partial{\mathcal H}
}{\partial{V}}\,,
\end{eqnarray}
where a dot denotes derivative with respect to the cosmic time $t$.
In the limit $\alpha\rightarrow\,0$, the Poisson algebra
(\ref{PA-IR}) and IR-deformed Hamilton's equations (\ref{EoM-IR})
and (\ref{EoM-IR2}) are reduced to their standard counterparts.
Indeed, depending on the value of the deformation parameter
$\alpha$, the system continue to follow the standard non-deformed
classical trajectories and deviations start to dominate for the
sufficiently large scales: $V\sim\alpha^{-1}$. Substituting the
Hamiltonian function (\ref{Hamiltonian}) into the relation
(\ref{EoM-IR}) and then using the Hamiltonian constraint
${\mathcal H}\approx0$, it is straightforward to show that the
IR-modified Friedmann equation will be
\begin{equation}\label{Friedmann0}
H_{_\pm}^2=\frac{\kappa}{3}\rho\left(1\pm\alpha^2V^2\right)\,.
\end{equation}
Differentiating the above relation with respect to time $t$ and
then using the energy conservation relation
\begin{equation}\label{energy-con0}
{\dot\rho}+3H_{_\pm}(\rho+p)=0\,,
\end{equation}
which is not modified in this setup (since just the geometric
parts are modified), one can easily obtain the following
IR-modified Raychaudhuri equation
\begin{equation}\label{Raychaudhuri0}
{\dot H}_{_\pm}=-\frac{\kappa}{2}\big[\rho\left(1\mp\alpha^2V^2
\right)+p\left(1\pm\alpha^2V^2\right)\big]\,.
\end{equation}
The IR effects would become significant at large volume limit at the late
time. We also expect that they would naturally address the late time cosmic
acceleration. In order to understand the late time behaviors of the above
IR-deformed models, we obtain the associated effective equation of state
parameters. Defining effective energy densities and pressures as
$\rho^{^\pm}_{_{\rm eff}}=\rho(1\pm\alpha^2V^2)$ and $p^{^\pm}_{_{\rm eff}}
=p\mp(2\rho-p)\alpha^2V^2$ through the relations (\ref{Friedmann0}) and
(\ref{Raychaudhuri0}), the effective equation of state parameters
$w^{^\pm}_{_{\rm eff}}=p^{^\pm}_{_{\rm eff}}/\rho^{^\pm}_{_{\rm eff}}$ can
be easily obtained as
\begin{eqnarray}\label{weff}
w^{^\pm}_{_{\rm eff}}=w\mp\frac{2\alpha^2V^2}{1\pm\alpha^2V^2}\,,
\end{eqnarray}
where $w=p/\rho$ is the standard equation of state parameter. The above
relation shows that $w^{^-}_{_{\rm eff}}\in[w,\infty)$ since $V\in[0,\alpha^{
-1})$. This general result show that the minus sign in (\ref{PA-IR}) cannot
generate the late time cosmic acceleration. Invoking the cosmological
observations which indicate that the universe accelerate at late time
\cite{DE}, we therefore abandon the minus sign. For the case of plus sign,
however, we have $w^{^+}_{_{\rm eff}}\in(w-2,w]$ when the volume changes as
$V\in[0,\infty)$. This is an interesting result since it shows that the plus
sign in (\ref{PA-IR}) can potentially address the dark energy problem
\cite{DEP,DEP2}. In the next section we show that the late time cosmic
acceleration naturally arises even in the cold dark matter dominated
universe.

\section{Dark Energy from Natural IR Cutoff}
We are interested in the late time cosmic evolution where the radiation
component is negligible. Also, we would like to address the dark energy
problem in the presented setup. Therefore, we consider the cold dark matter
(CDM) dominated universe without cosmological constant. In this respect,
our model has the same number of parameter as the standard $\Lambda$CDM
model such that the effects of cosmological constant will replace with IR
parameter $\alpha$.

\begin{figure}
\centering
\begin{minipage}{.49\textwidth}
  \centering
  \includegraphics[width=.9\linewidth]{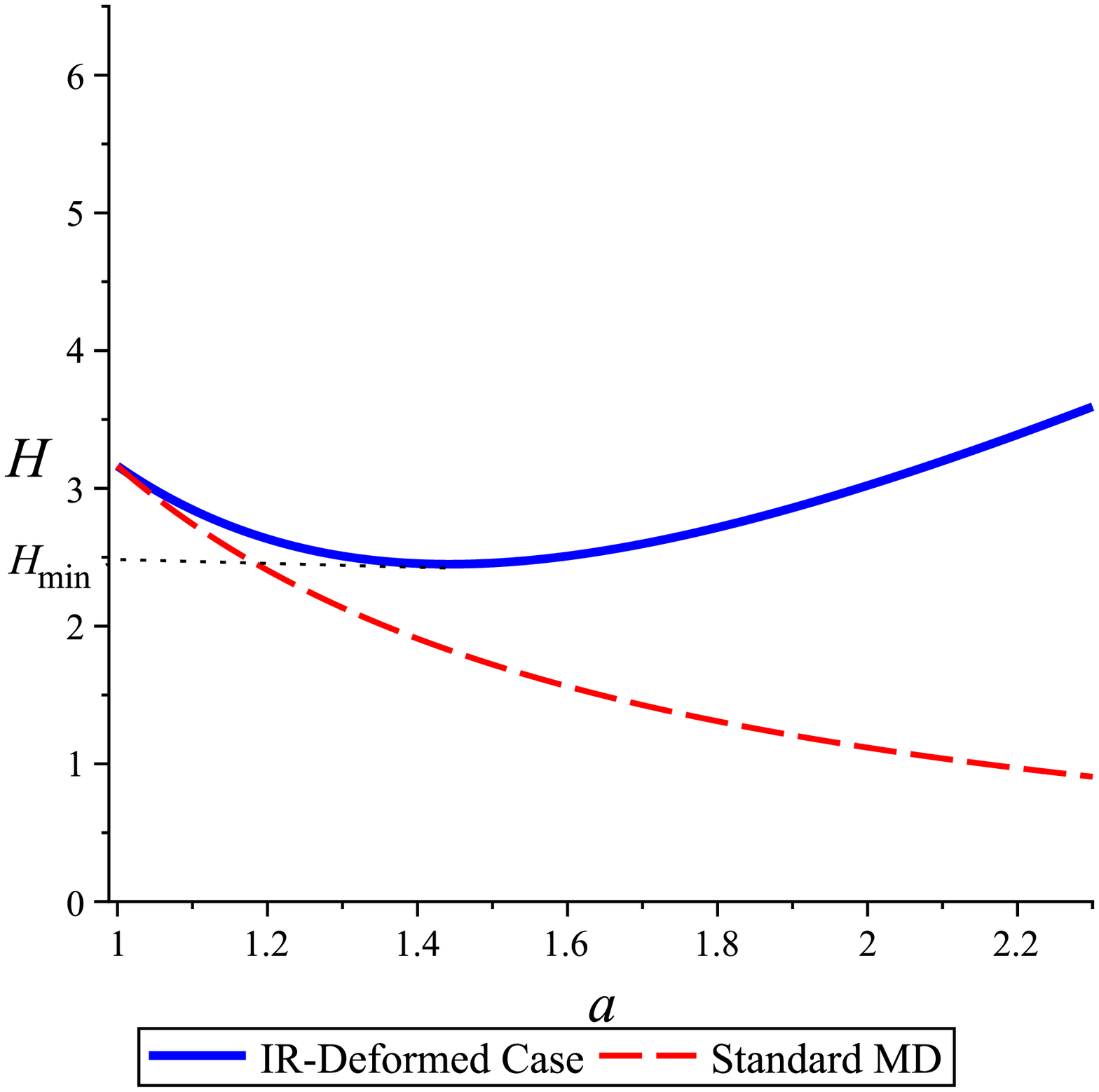}
  \caption{The Hubble parameter versus the
scale factor (as a clock) is plotted. While the Hubble parameter
decreases with decreasing rate in the standard matter dominated
era with $w=0$ (the red dashed line), it decreases with increasing
rate in our model until it approaches to its minimum
(\ref{hubble-minimum}) which is corresponding to an unstable de
Sitter phase with $w=-1$ (the blue solid line). After crossing
the phantom divide ($w<-1$) it starts to increase with increasing
rate and finally it diverges at the big rip with $w=-2$ through
the finite time (\ref{time-br}). The figure is plotted for
$\kappa=3$ and $\rho_{_{\min}}=1/3$.}
  \label{FIG1}
\end{minipage}%
\hfill%
\begin{minipage}{.49\textwidth}
  \centering
  \includegraphics[width=.9\linewidth]{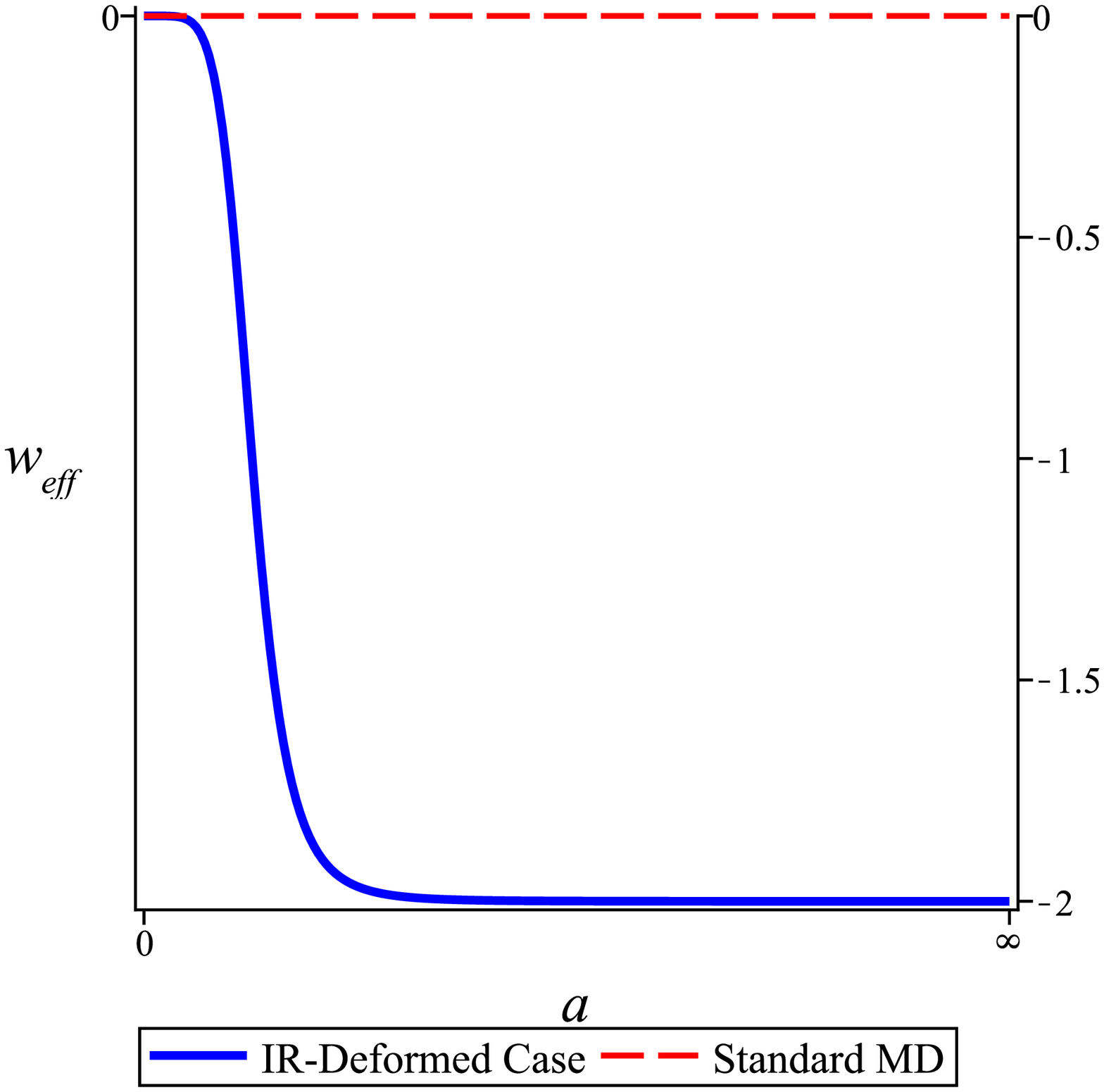}
  \caption{The effective equation of state
parameter versus the scale factor is plotted. It is clear that it
is bounded as $0\leq{w_{_{\rm eff}}}\leq2$ in this setup. The
unstable point $w_{_{\rm eff}}=0$ is corresponding to the matter
dominated era. As the universe expands, the IR effects starts to
dominate and after crossing an unstable de Sitter phase with
$w_{_{\rm eff}}=-1$, it enters in a phantom phase with
$w_{_{\rm eff}}<-1$. Finally, the universe ends up in a big rip
at the finite time (\ref{time-br}) with $w_{_{\rm eff}}=-2$. The
figure is plotted for $\rho_{_{\min}}=1$.}
  \label{FIG2}
\end{minipage}
\end{figure}

Considering the energy density $\rho_{_m}=\rho_{0_m}a_0^3a^{-3}$ for CDM in
(\ref{Friedmann0}) and (\ref{Raychaudhuri0}) (for the plus sign since we
have shown that the minus sign cannot produce acceleration), the IR-deformed
Friedmann and Raychaudhuri equations for the CDM dominated universe in this
setup are given by
\begin{equation}\label{Friedmann-IR}
H^2=\frac{\kappa}{3}\rho_{_m}\left(1+\frac{
\rho_{_{\min}}^2}{\rho_{_m}^2}\right),
\end{equation}
\begin{equation}\label{Raychaudhuri-IR}
\dot{H}=-\frac{\kappa}{2}\rho_{_m}\left(1-
\frac{\rho_{_{\min}}^2}{\rho_{_m}^2}\right),
\end{equation}
where we have defined the minimum energy density
\begin{equation}\label{rho-minimum}
{\rho}_{_{\rm min}}=\alpha\rho_{0_m}a_0^3\,.
\end{equation}
The above minimum energy density is defined in the sense that at $\rho_m={
\rho}_{_{\rm min}}$, the effective energy density
\begin{equation}\label{rho-total}
\rho_{_{\rm eff}}=\rho_{_m}+\frac{\rho_{_{\min}}^2}{\rho_{_m}},
\end{equation}
has the minimum $\rho_{_{\rm eff}}=2\rho_{_{\min}}$. The existence of this
minimum value for the effective energy density implies a minimum value for
the Hubble parameter through the relation (\ref{Friedmann-IR}) (see also the
figure \ref{FIG1}) which is given by
\begin{equation}\label{hubble-minimum}
H_{\rm min}=\sqrt{\frac{2\kappa\rho_{_{\min}}}{3}}.
\end{equation}

Before scrutinizing the cosmological implication of the model, it
is interesting to note that the IR-deformed Friedmann equation
(\ref{Friedmann-IR}) can be also deduced from the $f(R)$ theories
for the particular case of $f(R)=R+\alpha{R}^{-1}$ \cite{DE-FR2}.
Furthermore, it can be realized from the Cardassian models of dark
energy which is investigated in the context of braneworld scenario
\cite{cardassian}. In Cardassian model, the deformed Friedmann
equation in matter dominated era is given by $H^2=A\rho_m+B
\rho_m^n$ with $A=\kappa/3$ and $B$ and $n$ are two free
parameters of the model. Our model can be realized as a special
case by the relevant identification $B=\frac{\kappa}{3}
\rho_{_{\min}}^2$ and $n=-1$. Note that our model has also one
parameter less than the Cardassian models.

Solving (\ref{rho-total}) for $\rho_{_m}=\rho_{_m}(\rho)$ and then
substituting for the effective pressure $p_{_{\rm eff}}=p_{_m}+
p_{_d}=p_{_d}$ gives
\begin{equation}\label{p-rho}
p_{_{\rm eff}}=-\rho_{_{\rm eff}}\,\pm\,\rho_{_{\rm eff}}\,
\sqrt{1-4\Big(\frac{{\rho}_{_{\rm min}}}{
\rho_{_{\rm eff}}}\Big)^2}\,.
\end{equation}
The effective equation of state parameter is then given by
\begin{equation}\label{EoS}
w_{_{\rm eff}}=\frac{p_{_{\rm eff}}}{\rho_{_{\rm eff}}}=-1
\,\pm\,\sqrt{1-4\Big(\frac{{\rho}_{_{\rm min}
}}{\rho_{_{\rm eff}}}\Big)^2}\,.
\end{equation}
From the minimum bound that are arisen for the effective energy
density as $\rho_{_{\rm eff}}=2\rho_{_{\rm min}}$, it is clear
that the equation of state parameter (\ref{EoS}) is also bounded
as $-2\leq{w_{_{\rm eff}}}\leq0$ in complete agreement with our
pervious general treatment though the relation (\ref{weff})
(see figure \ref{FIG2}). This range for $w_{_{\rm eff}}$ shows
that the model can produce the acceleration phase.

To study the fate of the universe, using the IR-deformed
Friedmann equation (\ref{Friedmann-IR}) in the conservation
relation for the effective energy density
\begin{equation}\label{rho-con}
\dot{\rho_{_{\rm eff}}}+3H(\rho_{_{\rm eff}}+
p_{_{\rm eff}})=0\,,
\end{equation}
and then substituting for the pressure from the relation
(\ref{p-rho}) with minus sign (corresponding to the phantom phase)
gives
\begin{eqnarray}\label{time-br}
t_{_{\rm rip}}-t_{_{\Lambda}}=\frac{1}{\sqrt{3\kappa}}
\int_{2\rho_{_{\rm min}}}^{\infty}\frac{d\rho_{_{\rm eff}}
}{\sqrt{\rho_{_{\rm eff}}(\rho_{_{\rm eff}}^2-4{\rho
}_{_{\rm min}}^2)}}=\sigma\,H_{\rm min}^{-1}\,,
\end{eqnarray}
where $\sigma=\frac{2\sqrt{\pi}}{3}\frac{\Gamma[5/4]}{\Gamma[3/4]}
\approx0.874$ and also we have used (\ref{hubble-minimum}).
Clearly, the integration in (\ref{time-br}) is performed from the
lower bound $\rho_{_{\rm eff}}=2\rho_{_{\rm min}}$ corresponding
to the unstable de Sitter phase with $w_{_{\rm eff}}=-1$ (through
the relation (\ref{EoS})) to the final state with
$\rho_{_{\rm eff}}\rightarrow\infty$ and $w_{_{\rm eff}}=-2$.
Thus, $t_{_\Lambda}$ denotes the time at which the system is in
unstable de Sitter phase with effective cosmological constant
$\Lambda_{_{\rm IR}}=2\kappa\rho_{_{\min}}=2\kappa\alpha
\rho_{0_m}a_0^3$ and $t_{_{\rm rip}}$ corresponds to a big rip
since the effective energy density and Hubble parameter diverge
at finite time (\ref{time-br}) \cite{Caldwell}. To be more
precise, we should obtain the scale factor at the time of big rip.
Substituting pressure from the relation (\ref{p-rho}) with minus
sign into the relation (\ref{rho-con}) and integrating gives the
following integral for the scale factor at the big rip
\begin{equation}\label{sf-br}
a=\exp\left[\frac{1}{3}\int_{2\rho_{\rm min}}^{\infty}
\frac{d\rho_{_{\rm eff}}}{\sqrt{\rho_{_{\rm eff}}^2-4{
\rho}_{\rm min}^2}}\right]\longrightarrow\infty\,,
\end{equation}
which shows that it diverges for infinite energy density at the
big rip. From (\ref{p-rho}) it is clear that the pressure also
diverges at the time (\ref{time-br}) and therefore the universe
finally ends up in a big rip with $\rho_{_{\rm eff}},
|p_{_{\rm eff}}|,a\rightarrow\infty$ and $w_{_{\rm eff}}=-2$
at the finite time (\ref{time-br}). A big rip is the common
fate of the phantom dominated universe \cite{big-rip}. At
sufficiently high energy regime $\rho_{_{\rm eff}}\gg
\rho_{_{\min}}$, the IR effects are negligible and relation
(\ref{rho-total}) gives $\rho_{_{\rm eff}}\approx\rho_m$ which is
corresponding to the standard matter dominated era and therefore
the standard Friedmann and Raychaudhuri equations for the matter
dominated era can be recovered in this regime.

Note that fixing the IR deformation parameter of the model
$\alpha$ (or equivalently fixing $\rho_{_{\min}}$), which
replaces the cosmological constant in comparison with the
$\Lambda$CDM, immediately fixes all the observable parameters
such as the equation of state parameter $w_{_{\rm eff}}$ and
density parameters $\Omega_m$ and $\Omega_d$. Thus, fitting
the density parameter $\Omega_d$ with the
observational data immediately gives a fixed value for the
effective equation of state parameter $w_{_{\rm eff}}=
w_{_{\rm eff}}(\Omega_d)$ which shows the naturalness and
predictiveness of the model. Interestingly, the cases
$w_{_{\rm eff}}>-1$ and $w_{_{\rm eff}}<-1$ are corresponding to
$\Omega_m>\Omega_d$ and $\Omega_m<\Omega_d$ respectively which
shows the model is observationally viable. While the model
qualitatively is relevant, in contrast to the dynamical dark
energy models, it may not fit the observational data in a very
precise manner. But, note that the dynamical dark energy models
such as the model based on the scalar fields \cite{DE-SF} have at
least one parameter more that our model and $\Lambda$CDM. We
could therefore consider another model with more adjustable
parameters which fits the observational data in a more precise
manner. For instance, adding even a massless scalar field to the
matter content can produce cosmological constant like term at
late time in this setup. We are going to study such a setup in a
new research program \cite{IRDE-SF}. Moreover, similar to the
theories which deals with the geometric deformation of the
Einstein's equations at the late time such as $f(R)$ theories
\cite{DE-FR2,DE-FR} and recently proposed massive gravity models
\cite{DE-NIR}, our setup can produce an accelerating universe and
crossing the phantom divide without violating any energy condition.

\section{Autonomous System and Dynamical Stability}
In this section we consider the dynamical stability of the
self-accelerating CDM dominated universe that is presented
in the pervious section. We define the energy density and pressure
of the dark energy component as
\begin{equation}\label{rho-IR}
\rho_{_d}=\frac{\rho_{_{\min}}^2}{\rho_{_m}},
\end{equation}
\begin{equation}\label{p-IR}
p_{_d}=-2\rho_{_d}=-2\frac{\rho_{_{\min}}^2}{\rho_{_m}}.
\end{equation}
Using energy conservation relation ${\dot\rho}_{_m}+3H\rho_{_m}=
0$ for the pressureless CDM, it is easy to show that the energy
density (\ref{rho-IR}) and pressure (\ref{p-IR}) satisfy the
following conservation relation
\begin{equation}\label{energy-con}
{\dot\rho}_{_d}+3H(\rho_{_d}+p_{_d})=
{\dot\rho}_{_d}-3H\rho_{_d}=0\,.
\end{equation}
The IR-deformed Friedmann equation (\ref{Friedmann-IR}) then
rewrites as
\begin{equation}\label{Friedmann-IR2}
H^2=\frac{\kappa}{3}(\rho_{_m}+\rho_{_d}).
\end{equation}
From the above relation and the energy conservation relation
(\ref{energy-con}), one could easily find the IR-deformed
Raychaudhuri equation
\begin{equation}\label{Hubbledot-IR}
{\dot H}+H^2=-\frac{\kappa}{6}(\rho_{_m}
-5\rho_{_d})\,.
\end{equation}
As it is clear from the above relation, the energy density of the
dark energy $\rho_d$ which purely originates from the IR effects,
appears with the minus sign which show how it generates cosmic
acceleration at the late time.

In order to consider the dynamical stability of the model, we
work with the well-known dimensionless density parameters
\begin{eqnarray}\label{PD-IR}
\Omega_m=\frac{\kappa\rho_{_m}}{3H^2},\hspace{1cm}
\Omega_d=\frac{\kappa\rho_{_d}}{3H^2}\,,
\end{eqnarray}
in terms of which the IR-deformed Friedmann equation
(\ref{Friedmann-IR2}) becomes
\begin{equation}\label{Friedmann-DP}
\Omega_m+\Omega_d=1\,.
\end{equation}
From the definition (\ref{rho-IR}) and using the IR-deformed
Friedmann equation (\ref{Friedmann-IR}) together with the
definition of the effective energy density (\ref{rho-total})
it is easy to show that
\begin{equation}\label{DP-rho}
\Omega_m\Omega_d=\frac{H_{\rm min}^4}{H^4}
=\frac{\rho_{_{\rm min}}^2}{\rho_{_m}^2}\,,
\end{equation}
where we have also used the relations (\ref{rho-minimum}) and
(\ref{hubble-minimum}). This relation is useful to study the
qualitative behavior of the model.

Taking the time derivative of the dark energy density parameter as
${\dot\Omega}_d=\Omega_d\left(\frac{{\dot\rho}_{_d}}{\rho_{_d}}-
2\frac{\dot{H}}{H}\right)$, and then substituting from the
relations (\ref{energy-con}), (\ref{Friedmann-IR2}) and
(\ref{Hubbledot-IR}) gives
\begin{equation}\label{time-DP}
\frac{d\Omega_d}{d\tau}=6\Omega_d(1-\Omega_d)\,,
\end{equation}
where $\tau=\ln{a}$ and we have also eliminated the CDM density
parameter $\Omega_m$ by means of the constraint equation
(\ref{Friedmann-IR2}). As it is clear from (\ref{time-DP}), the
space of states is the one-dimensional
segment $\Omega_d\in[0,1]$ and there are two critical points
$\Omega_d=0$ and $\Omega_d=1$ in this model. The critical point
$\Omega_d=0$ clearly corresponds to the matter dominated era
with $\Omega_m=1$ and the point $\Omega_d=1$ is corresponding to
the dark energy dominated era with $\Omega_m=0$. In order to
consider the stability of the model, we should consider the
linear perturbation of the equation (\ref{time-DP}) around these
critical points. Considering the small perturbations $\Omega_d
=\delta_1\rightarrow\,0$ and $\Omega_d=1-\delta_2\rightarrow\,1$
in relation (\ref{time-DP}) immediately leads to the following
differential equations \cite{DS}
\begin{equation}\label{perturbation-DE}
\frac{d\delta_1}{d\tau}=6\delta_1,\hspace{1cm}
\frac{d\delta_2}{d\tau}=-6\delta_2,
\end{equation}
which have the following solutions
\begin{equation}\label{perturbation-DE-S}
\delta_1=c_1e^{6\tau},\hspace{1cm}
\delta_2=c_2e^{-6\tau},
\end{equation}
where $c_1$ and $c_2$ are constants of integrations. The point
$\Omega_d=0$ will be a past attractor since the initially small
perturbation $\delta_1$ increases exponentially with time and then
taking the system away from the matter dominated era $\Omega_m=1$.
The perturbation $\delta_2$, however, decreases exponentially with
the time which shows that the point $\Omega_d=1$ is a stable
equilibrium point. This is a future attractor solution which is
corresponding to a big rip at the time (\ref{time-br}) with $w=-2$.
While the total energy density behaves as $\rho\simeq\rho_{_m}$ at
the past attractor point $\Omega_d=0$ ($\Omega_m=1$ matter
dominated era) and the scale factor behaves as $a\propto\,t^{2/3}
$, it behaves as $\rho\propto\rho_{_m}^{-1}\propto{a^3}$ at the
future attractor point $\Omega_d=1$ ($\Omega_m=0$ dark energy
dominated era) and therefore the scale factor would behave as
$a\propto\,t^{-2/3}$.

\section{Summary and Conclusions}
Cosmological observations show that the universe accelerates at small redshifts and the 
cosmological constant derives the desired acceleration in standard $\Lambda$CDM 
cosmology. In the context of general relativity, the cosmological constant can be 
interpreted as a universal IR cutoff. Quantum gravity candidates such as loop quantum 
gravity and string theory also suggest the existence of a minimum length scale of the 
order of the Planck length. The ultimate quantum theory of gravity then should contain 
universal IR and UV cutoffs. While the standard general relativity accommodates the 
existence of IR cutoff through the de Sitter spacetime with positive cosmological
constant, it cannot support the existence of a UV cutoff. On the other hand, loop 
quantum cosmology scenario suggests the existence of a minimum length scale as UV cutoff 
for the system under consideration, but it does not support the existence of any IR 
cutoff. In this paper, following the UV deformation of the FLRW gravitational phase 
space in loop quantum cosmology, we have formulated deformed phase space which supports 
the existence of an IR cutoff. We obtained the associated IR-deformed Friedmann and 
Raychaudhuri equations and we showed that the IR corrections derives the late time 
cosmic acceleration. The model has the same number of parameters as $\Lambda$CDM such 
that the IR effects replace the effects of cosmological constant. But, the dynamics of 
the universe in our model is very different from the standard $\Lambda$CDM cosmology.
For instance the Hubble parameter and energy density turned out to be bounded from 
the below in this model. As a particular example, we applied the setup to the simple 
case of CDM dominated universe. We divide the cosmic evolution in this model into the 
following three phases:
\begin{itemize}
  \item {\it Quintessence phase} ($-1<w_{_{\rm eff}}\leq0$):
The universe starts from the standard matter dominated era with equation of state
parameter $w_{_{\rm eff}}\approx{w_m}=0$ for $\rho_{_{\rm eff}}\gg\rho_{_{\rm min}}$
when the IR effects are
negligible. In this phase, as the universe expands and the total energy density
(\ref{rho-total}) dilutes, the IR effects become more and more appreciable up to
$\rho_{_{\rm eff}}\sim\rho_{_{\rm min}}$. The effective equation of state parameter,
which is given by
the plus sign of the relation (\ref{EoS}) in this regime, then smoothly decreases
from $w_{_{\rm eff}}=0$ in matter dominated era to the negative values $w<0$ and
the universe starts to accelerate when it crosses over the value $w_{_{\rm eff}}=
-\frac{1}{3}$. From the relations (\ref{EoS}) and (\ref{DP-rho}), it is clear that
the accelerating phase with $-1<w_{_{\rm eff}}<-\frac{1}{3}$ is corresponding to
$\Omega_m>\Omega_d$.   \item {\it Unstable de Sitter phase} ($w_{_{\rm eff}}=-1$);
{\it Transition from} $w_{_{\rm eff}}>-1$ {\it to} $w_{_{\rm eff}}<-1$:
The effective equation of state parameter (\ref{EoS}) then approaches to the value
$w_{_{\rm eff}}=-1$ when the energy density of CDM approaches to the critical value
$\rho_{_m}=\rho_{_{\rm min}}$. At this momentum the Hubble parameter approaches to
its minimum $H=H_{\rm min}$ that is corresponding to an unstable de Sitter phase
with effective cosmological constant $\Lambda_{_{\rm IR}}=3H_{\rm min}^2=2\kappa
\alpha\rho_{0_m}a_0^3$. From (\ref{EoS}) and (\ref{DP-rho}), one can easily see
that the model includes a transition from the quintessence era with $w>-1$ and
$\Omega_m>\Omega_d$ to the phantom phase with $w_{_{\rm eff}}<-1$ and $\Omega_m<
\Omega_d$ when crossing over the unstable de Sitter phase with $w_{_{\rm eff}}=-1$
and $\Omega_m=\Omega_d$. This result makes the model observationally viable.
  \item {\it Phantom phase} ($-2\leq{w_{_{\rm eff}}}<-1$):
After the universe enters into a phantom era with $w_{_{\rm eff}}<-1$ and $\Omega_m
<\Omega_d$, the effective equation of state parameter decreases until approaches to
its asymptotic value $w_{_{\rm eff}}=-2$ where the universe ends up in a big rip
singularity at the finite time (\ref{time-br}).
\end{itemize}
We have also considered the dynamical stability of the model which shows that,
in this model, the universe starts at a past attractor in matter dominated era
($w_{_{\rm eff}}=0$) and after crossing an unstable point, corresponds to a
de Sitter phase ($w_{_{\rm eff}}=-1$), it approaches to a future attractor
($w_{_{\rm eff}}=-2$) which is corresponding to the big rip singularity.\\

{\bf Acknowledgement}\\
The author would like to thank Babak Vakili for a critical reading of the
manuscript, Kourosh Nozari for useful comments on the first draft of this work,
and Achim Kempf for helpful discussions. He also thanks anonymous referee for
very insightful comments which considerably improved the quality of the paper.

\end{document}